\documentclass[twocolumn,showpacs,preprintnumbers,amsmath,amssymb]{revtex4}
\usepackage{times} 
\usepackage{color}
\usepackage{graphicx}
\usepackage{ulem}
\usepackage{prelim2e}\usepackage[none,bottom]{draftcopy}
\draftcopyName{UT preprint / }{1.2}

\graphicspath{{.},{../Ps/},{figs/},%
{../figs/},%
{$LOCALHOME/Obj/Ratchet/Ps/}}
\newcommand{\mylab}[1]{\label{#1}}
\newcounter{countuwe}

\newcounter{countkar}

\newcounter{countfil}

\begin{document}
\title{Self-ratcheting Stokes drops driven by oblique vibrations}
\author{Karin John}
\email{kjohn@spectro.ujf-grenoble.fr}
\affiliation{Laboratoire de Spectrom\'etrie Physique UMR 5588, Universit\'e Joseph
Fourier - Grenoble I, BP 87 - 38402 Saint-Martin-d'H\`eres, France}
\author{Uwe Thiele}
\email{u.thiele@lboro.ac.uk}
\homepage{http://www.uwethiele.de}
\affiliation{Department of Mathematical Sciences, Loughborough
University, Loughborough, Leicestershire, LE11 3TU, UK}
\begin{abstract}
  We develop and analyze a minimal hydrodynamic model in the
  overdamped limit to understand why a drop climbs a smooth
  homogeneous incline that is harmonically vibrated at an angle
  different from the substrate normal [Brunet, Eggers and Deegan,
  Phys. Rev. Lett. \textbf{99}, 144501 (2007)]. We find that the
  vibration component orthogonal to the substrate induces a nonlinear
  (anharmonic) response in the drop shape.  This results in an
  asymmetric response of the drop to the parallel vibration and, in
  consequence, in the observed net motion.
  Beside establishing the basic mechanism, we identify
  scaling laws valid in a broad frequency range and a flow reversal at
  high frequencies.
\end{abstract}
\maketitle
%


The concept of transport driven by an external ratchet potential dates
back to Pierre Curie \cite{Curi1894}. He states that a locally
asymmetric but globally symmetric system may induce global transport
if it is kept out of equilibrium. Practical realizations include
colloidal particles that may move either through periodic asymmetric
micropores when driven by an imposed oscillating pressure field
\cite{MaMu03} or they can be driven by a sawtooth dielectric potential
that is periodically switched on and off \cite{RSAP94}.
Many different variations of ratchet mechanisms are nowadays studied
\cite{HaMa09}. They are employed to transport or filter
discrete objects \cite{RSAP94,MaMu03} or to induce macroscopic
transport of a continuous phase in systems without a macroscopic
gradient \cite{SISW03,QuAj06,BTS02}. Most modelling effort is
focused on the former, but first models do as well exist for the
latter \cite{Ajda00,JoTh07}.

An experiment that at first sight seems unrelated has recently shown
that drops may climb an inclined homogeneous substrate if it is
vibrated harmoniously in a vertical direction \cite{BED07}. 
The experiment is quite remarkable, as previously it had only been
shown that substrate vibrations can 'unlock' drops pinned by substrate
heterogeneities and therefore facilitate directed motion of drops in a
global gradient. In particular, they allow to overcome effects of
contact angle hysteresis on non-ideal substrates with a global
wettability gradient \cite{DaCh02,DSGC04}. A recent extension shows
that drops can as well be driven by simultaneous vertical and
horizontal substrate vibrations that are phase-shifted \cite{NKC09}.
Several hypothesis have been put forward as to why the vibrations
induce the drop motion. Contact angle hysteresis, nonlinear
friction, anharmonicity of the vibrations, convective
momentum transport are all mentioned as possible 'minimal ingredients'
\cite{BED07,NKC09,Beni09}.  Anharmonic lateral vibrations
with non-zero mean displacement are known to drive drop motion on
horizontal substrates \cite{DCG05}.
In general, several studies have adressed drops and free surface films
on oscillating substrates since early work by Faraday, Kelvin and
Rayleigh \cite{Lam32b}.  However, most consider fixed contact lines
\cite{StS84,CeK06}, purely inviscid flows \cite{LLS06,NBB04} or a high
frequency limit \cite{TVK06}; and, most importantly, limit their
study to vibrations either parallel or orthogonal to the substrate.

In the present Letter we analyze a minimal hydrodynamic model for the
situation depicted in Fig.~\ref{fig1} and show which of the above
mentioned ingredients are not necessary for the drop motion to occur.
\begin{figure}[tbh]
\includegraphics[width=0.6\hsize]{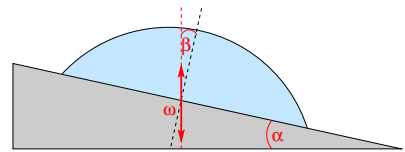}
\caption{(color online) Sketch of a drop on a vibrating (frequency
$\omega$, vibration angle with the substrate normal $\beta$) inclined
substrate (inclination angle $\alpha$).\mylab{fig1}}
\end{figure}
We derive a thin film evolution equation in the overdamped limit,
i.e., based on a pure Stokes flow (no convective momentum transport)
for a drop on an ideally smooth homogeneous inclined substrate (no
contact angle hysteresis).  The obtained results show that harmonic
vibrations are sufficient to drive a drop in a directed manner on a
horizontal substrate \cite{NKC09} or as well up an incline
\cite{BED07}.

Analysing the underlying mechanism we find that the component of the
harmonic vibration that is orthogonal to the substrate induces a
nonlinear (anharmonic) response in the drop shape.  As the latter
determines the strongly nonlinear drop mobility this results in an
asymmetric response of the drop to the vibration component that is
parallel to the substrate. The induced symmetry breaking between
forward and backward motion during the different phases of the
vibration results in the observed net motion of the drop.
This phenomenon can be seen as a rocked self-ratcheting of the drop
\cite{HaMa09} as it is the drop itself that introduces the local
time-reflection asymmetry in the response to the time-periodic driving
of the sliding motion.

%
%
In the following we first present the model, then analyze it
in the low frequency limit employing continuation techniques.  Results
for higher frequencies are obtained through numerical time integration.
We use a long-wave approximation \cite{ODB97,KaTh07} to describe the dynamics
of a drop of liquid on an inclined substrate (inclination angle
$\alpha$) that is subject to harmonic vibrations (frequency $\omega$,
at angle $\beta$ to the substrate normal, see Fig.~\ref{fig1}).  
We consider a two-dimensional (2d) drop as we do not
expect any conceptual difference to the 3d case.
%
The resulting evolution equation for the film thickness profile $h(x,t)$ is
\begin{equation}
\partial_t\,h\,=\,\partial_x\,\left[\frac{h^3}{3\eta}
\left(\partial_x\,p 
- f\right)\right].
\mylab{film}
\end{equation}
The divergence of the flow on the r.h.s.~is expressed as
the product of a mobility and the sum of a pressure gradient
$\partial_x p$ and a lateral driving force $f$. $\eta$ is the dynamic
viscosity. 
%
The pressure
\begin{equation}
p =-\,\gamma\partial_{xx} h\,-\,\Pi(h) + \rho g h\left[1+a(t)\right]
\mylab{press}
\end{equation}
contains the curvature pressure $-\gamma\partial_{xx} h$, where
$\gamma$ denotes surface tension, the disjoining pressure $\Pi(h)$
that incorporates wettability \cite{deGe85,ODB97} and the hydrostatic
pressure $\rho g h[1+a(t)]$, where the time dependence results from
the vibration component normal to the substrate.
The lateral force 
\begin{equation}
f=\rho g \left[\alpha+\beta a(t)\right]\,.\mylab{forceeq}
\end{equation}
contains a constant part (force down the incline) and a time-modulated
part (vibration component parallel to the substrate).  The function
$a(t)=a_0\,\sin(\omega t)$ corresponds to the acceleration in units of
$g$. The drop experiences a force of opposite sign, e.g., if the
substrate is accelerated upwards and to the left the drop is flattened
and pushed to the right.
%
%
The partial wettability of the substrate is modelled through a
precursor film model based on the disjoining pressure
$\Pi=\left(A/h^3+B/h^6\right)/6\pi$.
It combines long-range destabilizing ($A<0$) and short-range
stabilizing ($B>0$) van der Waals interactions \cite{Pism01}.

Note, that consistency of the long-wave approach requires small free
surface slopes as well as $\alpha,|\beta|\ll1$ \cite{KaTh07}.
However, it is known that equations like (\ref{film}) often predict
the correct qualitative behavior even for systems with larger contact
angles \cite{ODB97}. In consequence, we expect the here obtained
results to hold qualitatively as well for larger $\beta$.

To nondimensionalize we introduce the scales
$t_0=3\gamma\eta/h_0\kappa^2$, $x_0=\sqrt{\gamma h_0/\kappa}$, and
$h_0=(B/A)^{1/3}$ for $t$, $x$, and $h$, respectively, where $\kappa=A/6\pi h_0^3$.
The resulting dimensionless equations correspond to Eqs.\
(\ref{film})-(\ref{forceeq}) with $3\eta=\gamma=B=1$ and $A=-1$. 
The dimensionless modulated hydrostatic pressure and lateral force are
$p_h=Gh\,[1+a(t)]$ and $f=G(\alpha+\beta a(t))$ with $G=\rho g
h_0/\kappa$. 
The vibration
period $T=2\pi/\omega$ is given in units of $t_0$.
The fixed drop volume
$V=L(\bar h-h_p)$ is determined by the domain size $L$, the mean film
thickness $\bar h$ and the precursor film thickness $h_p=1$.  The
resulting transport along the substrate is measured after all
transients have decayed and the vibration-induced shape changes of the
drop are completely periodic in time. We quantify the transport by the
mean drop velocity $\langle v \rangle=\Delta x/T$ where $\Delta x$
denotes the distance the drop moves within one period $T$.

Fig.\,\ref{evol1} illustrates the typical behaviour of a drop during
one vibration cycle. One notices forward/backward lateral motion and
small but significant changes of shape.
\begin{figure}[tbh]
\includegraphics[width=0.7\hsize]{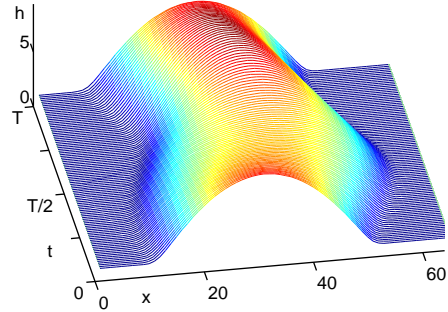}
\caption{(color online) Space-time plot illustrating the evolution of
  the profile of a drop on a obliquely vibrated substrate during one
  vibration period. Each cycle results in the net motion of the drop
  to the left. The starting time is well after initial
  transients have decayed. Note that only part of the domain $L$ is shown. 
Parameters are $V=192$, $G=0.001$, $\beta=0.1$, $\alpha=0$,
  $T=400$, $L=128$, $a_0=10$.} \mylab{evol1}
\end{figure}
In particular, the drop becomes flatter and broader during the first
  part of the cycle when its center of mass moves to the right
  ($t<T/2$, substrate acceleration is upward and to the left.  In the second half of
  the cycle the drop becomes again higher and less wide while it moves
  to the left ($t>T/2$, substrate acceleration is downward and to the
  right). After one period the drop has moved a net distance to the
  left.  In the course of one period, the orthogonal component of the
  oblique vibration modulates the hydrostatic pressure and causes a
  nonlinear response in the drop shape. That in turn determines the
  strongly nonlinear drop mobility [$h^3/3\eta$ in Eq.~(\ref{film})]
  and therefore induces an anharmonic response of the drop to the
  harmonic parallel vibration component that results in the observed
  net motion. Note, that a harmonic back and forth forcing alone
  results in zero net motion.

First, we consider a slowly vibrating substrate, i.e.\
the typical timescale of the intrinsic drop dynamics $t_0$ is small
compared to the vibration period $T$.
In this low frequency limit the drop moves in a quasi-stationary
manner, i.e., its shape and velocity at each instant during the
vibration cycle are those
of a stationary moving drop at the corresponding acceleration.  The resulting
family of moving drop solutions is parametrized by $a(t)$ and can be
obtained using continuation techniques \cite{DKK91,DKK91b,Thie02}. Averaging
stationary drop velocities $v(a(t))$ over one vibration period gives
the low frequency limit of $\langle v\rangle$.

\begin{figure}
\includegraphics[width=0.7\hsize]{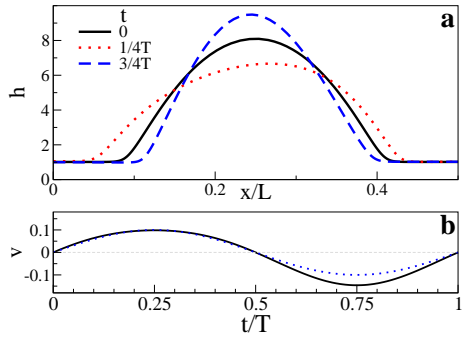}
\caption{(color online) Shown are (a) several drop shapes and (b) the
  velocity during one vibration cycle for an obliquely vibrated
  horizontal substrate in the low frequency limit.  For the used
  $\beta=0.3$ the drop moves with $\bar v \approx -0.01$, i.e., to the
  left. The blue dotted line in (b) indicates a harmonic variation of
  zero net flow.  Remaining parameters are as in Fig.~\ref{evol1}.
  \mylab{dropforms}}
\end{figure}

\begin{figure}
\includegraphics[width=0.7\hsize]{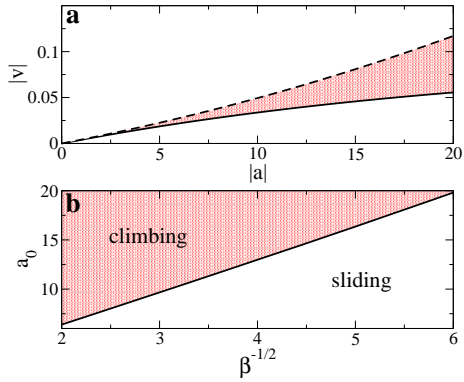}
\caption{(color online) (a) Mean drop velocity depending on the
acceleration $a$ for a horizontal substrate. The solid (dashed) line
indicates $a>0, v>0$ ($a<0, v<0$).
(b) Phase diagram indicating the regions of sliding and climbing
drops in the $a_0-\beta^{-1/2}$-plane for a substrate inclination of
$\alpha=0.1$.
The results for (a) and (b) are in the low frequency limit with the
remaining parameters as in Fig.~\ref{evol1}.
\mylab{staddrop}
\mylab{va0}}
\end{figure}

Figs.\,\ref{dropforms}\ (a) and (b) show profiles of stationary moving
drops and their velocities at various phases of the cycle on a
horizontal substrate.  As observed already in Fig.~\ref{evol1}, during
the first [second] half cycle the drop is compressed [decompressed]
and slides to the right [left].
Fig.\ \ref{va0} shows the relationship between acceleration $a$ and
the drop velocity $v$.  The difference between the two respective
curves for positive and negative acceleration is a measure of the
anharmonic response of the drop.  
The mean
velocity in the low frequency limit then corresponds to twice the
weighted area of the shaded region in Fig.\ \ref{va0}(a) when taken
from $a=0$ to $a=a_0$. 
A numerical analysis gives $\langle v\rangle\sim a_0^2$,
$\langle v\rangle\sim \beta$ and $\langle v \rangle\sim V^{1.67}$.
Note that for positive $\beta$ in the present limit net transport is
always directed to the left.

From the transport properties on a horizontal substrate it is
obvious, that such a vibration-caused motion can overcome a further
external driving and, e.g., move a drop up an incline or against a
wettability gradient, as long as the product $\beta a_0^2$ is larger
than a critical value. 
%
This is confirmed by the phase diagram Fig.\,\ref{staddrop}(b), as
there the straight line $a_0
\sim \sqrt{v_0/\beta}$ separates climbing from sliding drops, where
$v_0$ is the drop velocity for the given substrate inclination without
any vibration.  Note that the case $\beta=\alpha$, i.e., a vibration
vertical in the laboratory frame, is by no means special: uphill
motion is generated above a finite threshold acceleration. However, in
the limit $\beta\rightarrow 0$, i.e., for a vibration normal to the
substrate, the threshold acceleration becomes infinite and the net
transport is zero.

\begin{figure}
\includegraphics[width=0.7\hsize]{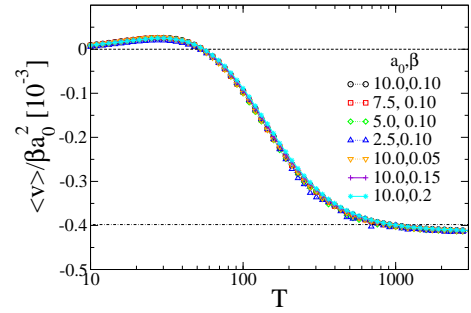}
\caption{(color online)
  Master curve for the scaled mean velocity $\langle v\rangle/\beta
  a_0^2$ as a function of the vibration period $T$ for sets ($a_0$,
  $\beta$) as given in the legend. The horizontal dashed line
  indicates the result in the low frequency limit.  Remaining
  parameters are as in Fig.~\ref{evol1}.  \mylab{flT} }
\end{figure}

For practical purposes, e.g., for microfluidic applications, drops
need to be transported in a limited time, i.e., the behaviour at
sufficiently large frequencies has to be understood.  We study that
regime employing an adaptive time-step 4th order Runge-Kutta method
and prevent numerical instabilities by a switching upwind differencing
for the driving term.
The above deduced scaling
$\langle v\rangle \sim a_0^2\beta$ still holds in a large frequency
range, illustrated in Fig.\,\ref{flT} by collapsing dependencies of
mean velocity on vibration period for various sets ($a_0$, $\beta$)
onto a single master curve.
%
A second interesting feature is the
observed reversal of net transport at a small critical period $T_{\rm
c}\approx55$ that results in a (small) positive mean velocity even for
a horizontal substrate ($\alpha=0$ as in Fig.\,\ref{flT}). The
reversed flow is strongest at $T_{\rm max}\approx T_{\rm c}/2
\approx27$, but always about one order of magnitude smaller than the
transport to the left in the low frequency regime.  Note that in
agreement with the first observation $T_{\rm c}$ and $T_{\rm max}$ are
nearly independent of $a_0$ and $\beta$.
\begin{figure}
\includegraphics[width=0.7\hsize]{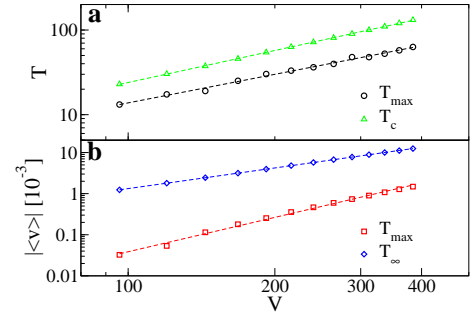}
\caption{(color online) Dependence of transport on drop volume: (a)
shows the special vibration periods $T_{\rm max}\sim V^{1.12\pm 0.03}$
and $T_{\rm c}\sim V^{1.26}$, as detailed in the text; (b) gives the
absolute value of the mean velocities $\langle v \rangle _{\rm max}
\sim V^{2.78\pm 0.06}$ at period $T_{\rm max}$ and in the low freqency
regime ($T_\infty$). Remaining parameters are as in Fig.~\ref{evol1}.
  \mylab{TcritL}}
\end{figure}
They do, however, depend on drop volume (see Fig.\ \ref{TcritL})
indicating that the flow reversal might be triggered when the
vibration frequency becomes larger than an eigen frequency of the
drop, i.e., for larger imposed frequencies the response of the drop
becomes delayed and phase-shifted w.r.t.~the forcing. This is
corrobated by an inspection of drop profiles over the course of a
cycle (not shown). This result is similar to the dependence of the
direction of net motion on the phase-shift between vertical and
horizontal vibration observed in \cite{NKC09}.
\begin{figure}
\includegraphics[width=0.7\hsize]{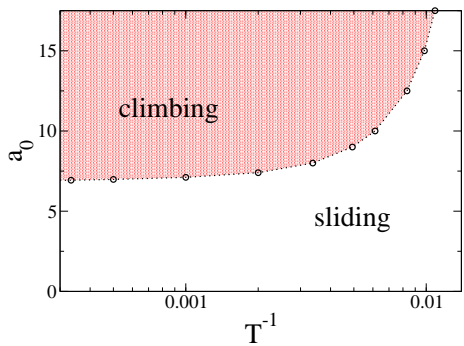}
\caption{(color online) Phase diagram of sliding and climbing drops in
  the $(a_0, 1/T)$ plane for an obliquely vibrated inclined
  substrate. Beside $\alpha=0.05$ parameters are as in
  Fig.~\ref{evol1}.  \mylab{Tca0alpha05}}
\end{figure}

Finally, Fig.\ \ref{Tca0alpha05} presents a phase diagram that shows
the influence of the vibration period and peak acceleration on the
transport direction for fixed substrate inclination angle
$\alpha=0.05$. The critical acceleration $a_{\rm c}(T^{-1})$ separating
sliding and climbing drops approaches a constant at small frequencies
and diverges at finite $T_{\rm c}^{-1}$. As we do not model any effects 
of non-ideal substrates (contact angle hysteresis, contact line pinning)
our phase diagram differs in two aspects from the experimental one
\cite{BED07}: (i) we find a monotonic
decrease of $a_{\rm c}$ with decreasing frequency ($T^{-1}$) whereas
\cite{BED07} finds evidence for a small increase at small $T^{-1}$;
and (ii) here drops always slide down below $a_{\rm c}$ whereas in the
experiments one finds instead a transition between static and climbing
drops at large $T^{-1}$.

To conclude, we have proposed and analyzed a minimal hydrodynamic
model for the experimentally observed drop motion that is driven by
harmonic oblique substrate vibrations \cite{BED07}.  Our analysis has
ruled out convective momentum transport, anharmonicity of vibrations
and contact angle hysteresis as necessary for the motion.
The mechanism that moves drops on a horizontal substrate or even up an
  incline is based on a nonlinear response of the drop shape to the
  normal vibration component. This breaks the back-forth symmetry of
  the response of the drop to lateral oscillations and therefore
  causes a net motion. The found mechanism is in line with the
  hypothesis put forward in \cite{BED07} based on a mechanical
  analogue that a breaking of the front-back symmetry due to a
  nonlinear friction law is sufficient to induce transport. Here, we
  have identified the strongly shape-dependent nonlinear mobility in
  Eq.~(\ref{film}) as the relevant hydrodynamic 'nonlinear
  friction'. Beside establishing the basic mechanism, our analysis has
  revealed several interesting features that should be investigated in
  future experiments. In particular, we have identified a number of
  scaling laws valid in a broad frequency range and found a flow
  reversal at high frequencies.

We have argued that the phenomenon can be seen as a self-ratcheting of
  the drop as its shape changes are instrumental in producing local
  time-reflection asymmetries. Comparing Eq.~(\ref{film}) with
  Fokker-Planck descriptions for ratchet systems of interacting
  particles \cite{HaMa09,SMN04} one may further the analogy by noting
  that (i) the lateral vibration component corresponds to an imposed
  rocking, (ii) the orthogonal vibration component corresponds to an
  imposed temporal temperature modulation, (iii) the role of the
  spatial asymmetry in the ratchet potential is here taken by
  non-linear couplings due to strongly nonlinear prefactors of the
  diffusion (2nd order) and the transport (1st order), (iv) our
  surface tension term is analogous to a mean field expansion of the
  distribution function up to second order.
%
%
%



%
%
\end{document}